# Laser spectroscopy with nanometric gas cells : distance dependence of atom-surface interaction and collisions under confinement

I.Hamdi, P.Todorov, A.Yarovitski, G.Dutier, I.Maurin, S. Saltiel*, Y.Li**,
A.Lezama[#],T.Varzhapetyan[(§)], D.Sarkisyan[(§)], M.-P.Gorza, M.Fichet, D.Bloch, M.Ducloy

*Laboratoire de Physique des Lasers, UMR 7538 du CNRS et de l'Université Paris13*
*99 Av JB Clément, F-93430 Villetaneuse, France*
*e-mail: ducloy@lpl.univ-paris13.fr*

[(§)] *Institute for Physical Research, Armenian Academy of Sciences, Ashtarak2, Armenia*

* *Permanent address: Department of Physics, University of Sofia, Sofia, Bulgaria*
** *Now at: Department of Physics, Guyuan Normal University, Ningxia, China*
[#] *Permanent address: IFFI, Universita de la Republica, Montevideo, Uruguay*

**Abstract** *The high sensitivity of Laser Spectroscopy has made possible the exploration of atomic resonances in newly designed "nanometric" gas cells, whose local thickness varies from 20nm to more than 1000 nm. Following the initial observation of the optical analogous of the coherent Dicke microwave narrowing, the newest prospects include the exploration of long–range atom surface van der Waals interaction with spatial resolution in an unprecedented range of distances, modification of atom dielectric resonant coupling under the influence of the coupling between the two neighbouring dielectric media, and even the possible modification of interatomic collisions processes under the effect of confinement.*

The high sensitivity of Laser Spectroscopy has recently made possible the exploration of atomic resonances in newly designed "nanometric" gas cells. We discuss here some of the novel features and new prospects that arise from these resonant confined gases.

**1. Optical spectroscopy in micro- and nano-cells**

For a very dilute gas, it is rather common, although an unfamiliar idea, that the atomic "mean free path" is actually governed by the container geometry. Designing specific vapour cells, with parallel windows and a very small thickness, enables one to make the "mean" free path strongly anisotropic, as atoms fly from wall to wall if the vapour is dilute enough. In these conditions, the transient build-up of the resonant interaction with light is responsible for a specific enhancement of the response of the "slow" atoms (with respect to the normal velocity, *i.e.* the velocity component along the normal to the windows). As long as the irradiation is under normal incidence, this had provided the principle of a novel method for Doppler-free spectroscopy, applicable to a variety of situations (velocity-dependent optical pumping, linear absorption, two-photon transition, *etc.* ; see [1] and refs. therein). The initial demonstrations were operated with glass cells of commercial origin, filled-up with alkali-metal vapours, and whose thickness (10-1000 μm) remained much larger than the optical wavelength λ. Because of that, the amplitude of the specific sub-Doppler signal was limited, and the effects of the atom-surface interaction [2] had remained unobservable, while we could currently investigate them (see *e.g.* [3] and refs. therein) through developments in selective reflection (SR) spectroscopy, a method known to offer a probing on a typical reduced wavelength λ/2π ~100 nm.

The recent fabrication of extremely thin cells (ETC) of vapour [4], designed with an initial sub-micrometric spacer between two carefully polished parallel windows, has revealed that under the effect of the external atmospheric pressure, the local cell thickness can become extremely small, typically spanning from 20 nm to 1μm. This truly permits to investigate "nanocells" of vapour, and has opened a realm of novel prospects, that even extends beyond sub-Doppler spectroscopy, with the additional possibility of detecting atom-surface interaction effects in an unexplored range of distances.

The intrinsic accurate parallelism of the windows (*e.g.* deviation < 1μm over a 10mm transverse extension), implies an intrinsic Fabry-Perot behaviour. Although convenient to evaluate the local thickness of the nanocell through an irradiation at nonresonant wavelengths, this behaviour tends to unavoidably mix-up the elementary behaviours associated to reflection and transmission spectroscopies [5]. Also, it naturally exhibits a λ/2 periodicity.

Aside from this periodic interference effect, the spectral width exhibits complex variations with the nanocell length, as a combined result of the simultaneous narrowing associated to the slow atoms contribution, and of the limited interaction time, that implies finally a broadening for the smallest thickness. More surprisingly, one also observes a λ-pseudo-periodicity in the resonant linear atomic response, with the sharpest spectrum obtained for a λ/2 thickness [6]. Indeed, the linear atomic response features a *coherent* spectroscopic narrowing, that extends, to the optical domain, the original observation by Romer and Dicke [7] in the microwave domain, of a *coherent* spectroscopic narrowing in a gas sample with a thickness ~ λ/2. Additionnally, a periodical revival of the Dicke-narrowing [6,8] was demonstrated (see fig.1). This kind of coherent Dicke narrowing can be viewed as one aspect of the more general consideration, by Dicke [9], of the effects of sub-wavelength confinement in atomic physics, with his related analysis of the Doppler effect in the time domain, rather than in the more usual frequency-domain.

The coherent spectral narrowing [6-8] relies on the fact that, when an atom leaves the wall, the sudden atom excitation induced by *on-resonance* light starts to precess in phase with the electromagnetic field at the wall position, but falls out-of-phase with the local driving field under the effect of the atomic motion, with a phase mismatch finally reaching kL (L: the cell thickness, k the wave number). As shown in a Bloch vector model [8], the phase mismatch appearing on line centre under a weak driving field turns to be independent of the atomic velocity because, for wall-to-wall trajectories, it accumulates during a duration L/v, at the rate of the Doppler shift kv (v being the normal atomic velocity). Hence, for a cell length up to $\lambda/2$ (*i.e.* maximal phase shift : $\pi$), all regions of the cell interfere constructively, whatever the velocity may be, yielding for L=$\lambda/2$ a strong contribution at line-centre. Oppositely, the signal tends to vanish for L=$\lambda$, as a result of an overall destructive interference between the different regions. For a frequency-detuned irradiation, the overall angular precession of the atomic excitation becomes velocity-dependent, and leads to a smooth length dependence of the signal that justifies the contrast between the line centre and the wings for L=$\lambda/2$. One even observes, for L=$3\lambda/2$, a revival of the sharp peak at line-center, although its amplitude, relatively to the background, decreases. This oscillating behaviour, with a $\lambda$ (pseudo-) periodicity of the coherent narrowing, fades away for larger values of L, as the atomic dipole relaxation has to be considered when it occurs on a time scale shorter than the time-of-flight between the walls (*i.e.* the dipole mean free path gets small compared to the cell thickness L). The spectral consequences of this behaviour are illustrated in fig. 2**.**

A fluorescence detection, easily implemented for atomic transitions excited in the visible or near-infrared range [4,8], also exhibits a strongly sub-Doppler excitation lineshape for short cell lengths. However, the simple picture of a non-degenerate two-level atom, enabling again to use the Bloch vector model, shows [8] that the fluorescence is essentially unsensitive to the interferometric effects that are responsible for the Dicke narrowing. This difference with the linear coherent transmission can be understood in a more general manner from the incoherent nature of the fluorescence process. In addition, the fluorescence, being a second-order process, is limited to those atoms slow enough to undergo a successive absorption and emission process. This makes its lineshape narrower than the transmission profile, even at its narrowest, *i.e.* for the L= λ/2 situation.

The linewidth of fluorescence is actually comparable to the one obtained for FM transmission. Indeed, the FM version of nanocell spectroscopy technique is known to be particularly appealing, as for all other *linear* Doppler-broadened systems sensitive to transient effects, such as SR spectroscopy [2,3]. With its use of a synchronous detection of the optical signal relatively to an applied Frequency Modulation, the FM technique provides indeed the frequency-derivative of a direct lineshape characterized by a central logarithmic spectral singularity. This singularity, typical of thin cell spectroscopy, is turned into a pure Doppler-free lineshape through the FM technique. In the elementary case of a λ/2 cell, the usual sub-Doppler linear signal is observed to be narrowed down to a pure Doppler-free signal, while oppositely, the peak-amplitude of the measured signal is not reduced, at least as long as the FM amplitude is comparable to the optical width. This makes the nanocell transmission spectroscopy one of the very few spectroscopic methods that can be operated in a vapour while being both linear and Doppler-free. It is planned to become an alternate method to saturation spectroscopy for weak, hard-to-saturate, transitions, notably to obtain frequency references from weak molecular transitions that are not suitable for saturated absorption spectroscopy.

Saturation effects in nanocells are also under investigation [10]. The overall trend can be viewed as a survival of the narrow resonances at large intensities, in spite of a washing out of the rapid length dependence (in transmission-reflection experiments) associated to the coherent Dicke narrowing. When optical pumping to an adjacent energy level can occur, the saturation effects, early shown to be velocity-selective and responsible for sub-Doppler features in the transmission for a thin cell [11], are incoherent, and like for fluorescence, are unsensitive to the coherent Dicke narrowing. This implies that the residual oscillating length dependence, experimentally observed in transmission on an *open* two-level systems under strong irradiation, should be only traced back to a first-order effect. A systematic study of saturation effects with nanocells exhibit complex saturation behaviours, including sharp length dependence, varying with the considered different hyperfine component. For the cycling transitions of a degenerate two-level system (*i.e. close* two-level system), when the associated coherent behaviour is susceptible (in the free-space) to lead either to a dark absorption line (J→J'=J-1), or to a bright absorption (J→J'=J+1), one can observe, over the broad resonant behaviour, dips or inverted dips at line centre depending on the cell length, in a manner reminiscent of the coherent Dicke narrowing. Theoretically, for these pure 2-level system, the situation is complex because saturation effects remain in the frame of a strong but coherent excitation, so that one may anticipate a residual coherent Dicke effect. However, thin cell spectroscopy intrinsically explores the dynamics of the atomic response, with a velocity distribution that scrambles this dynamics by the mixing of various time scales, owing to the different Rabi frequencies involved with the Zeeman sub-states. This complexity of the time sequence of the nonlinear process at least hinders any recognizable Dicke-type pseudo-periodicity.

**2. Exploring the atom-surface van der Waals interaction at small distances**

As mentioned in section 1, the local thickness of the cell can be monitored through careful interference measurements (*e.g.* in reflection) operated at various non resonant wavelengths. On the Cs resonance line (*e.g.* $D_1$ line, λ=894 nm), no notable deviations were observed relatively to the predictions for a standard isolated 2-level atom down to a λ/4 thickness [6]: the variations of the spectral lineshapes with the cell length -in transmission and in reflection as well - simply result from the coherent narrowing, combined with the Fabry-Perot interferences. Indeed, the long-range atom-surface interaction is dominated by the universal van der Waals (vW) attraction $V(z) = - C_3 z^{-3}$ (z: the atom-surface distance) between a fluctuating atom dipole and its image. This attractive potential, whose strength grows with the atomic excitation, induces a spectral red-shift on the $D_1$ transition ~ 1.2 MHz @ 100 nm (for an interaction with a single YAG wall), that is too small to be observed in nanocell spectroscopy for relatively large thickness. With respect to the Doppler width ( ~250 MHz) and to the natural transition width (~5 MHz), the onset of a practical observation of the vW interaction appears only for a cell thickness below 200 nm -*i.e.* all atoms are at a distance from the wall closer than 100nm -.

The major interest of these nanocells for the probing of the vW interaction is that it extends the possibility to probe short-lived excited states, typical of optical methods such as SR spectroscopy, to an unusual range of short distances. This range can normally be arbitrarily chosen through the choice of the container thickness instead of being imposed by the wavelength of observation -as it is the case in SR spectroscopy, with a probed depth ~ $\lambda/2\pi$ - . This interest is enhanced by the fact that the choice of methods to investigate the long-range atom-surface interaction is actually scarce, in spite of the ubiquity of the vW interaction. Most often, these methods rely on mechanical effects -even for laser cooled atoms- and are applicable only to ground state atoms or long-lived states. This explains that the only accurate measurements of the distance law of the interaction have been limited to the 3000-500 nm range (with Rydberg atoms that are strongly interacting and long-lived, allowing for a detection that is partly mechanical) [12]. With nanocells, one envisions the possibility of probing distances at least an order of magnitude smaller, reaching a distance range where limitations to the $z^{-3}$ scaling could originate in various short range effects, not always precisely investigated, such as the details in the cell roughness, or interaction with impurities sticking onto the surface. This $z^{-3}$ scaling normally applies in a typical 1000-1 nm range [2], *i.e.* as long as retardation (Casimir) effects are negligible and before occurence of the short range effects unavoidably related to a dependence on the atomic structural details of the surface. It hence implies a validity covering a huge energy range, that is particularly worth to be tested.

Various types of experiments on nanocells have evidenced the effect of the atom-surface interaction. From our most recent analyses, we now expect to be able to shift from the simple observation [13] of the surface interaction, to its effective measurement. In particular, it would be very interesting to evaluate the possible accuracy of a vW measurement in nanocell spectroscopy, with respect to that provided by well-known SR spectroscopy.

Our first step was to analyze the transmission behaviour in nanocells on the Cs $D_1$ resonance line. When the local cell thickness turns to be very small (~ 50-100nm), we observe a lineshape distortion and red-shift (up to 200MHz) that largely exceeds those observed through SR spectroscopy. Moreover, the spectral lineshapes appear in good agreement with a theoretical model that assumes a known strength of the van der Waals (vW) interaction. This theory simply assumes a thermal distribution of atoms flying wall-to-wall, and integrates the transient atomic response while taking into account a vW potential. The vW potential itself is modelled according to an electrostatic description including the interaction with the multiple images induced in the two reflecting walls. These transmission experiments have been extended to experiments in the FM mode, whose improved spectral resolution is of particular interest for large thickness (*i.e.* when the vW shift remains small), and to reflection spectra : for the smallest thicknesses, reflection spectra are recorded over a low level of non resonant reflection as due to FP interferences, and offer a competitive possibility to observe the vW shift [13]. More recently, and as discussed in more details in section 4, attempts to fit the recorded spectra with adjustable theoretical curves reveals that taking into account the vW interaction slightly improves the fitting for a 223 nm (~$\lambda/4$) thickness (see fig.3), while no improvement appears for a $\lambda/2$ thickness.

We are also investigating the stronger vW shift induced on high-lying excited states, such as Cs(6D), that is probed at 917nm ($6D_{5/2}$) or 921 nm ($6D_{3/2}$) after a prior excitation on the Cs $D_2$ line $6S_{1/2} \rightarrow 6P_{3/2}$ at 852 nm (see fig.4). For these 917 nm and 921 nm transitions, the vW interaction should be about an order of magnitude larger than for the $D_1$ line, at least in the absence of a resonant coupling between the atom excitation and a surface mode. Thanks to a laser diode specially designed for a broad range tuneability (through step-by-step frequency changes), we could observe the broadened transmission (and reflection) spectra for various thicknesses, down to 20 nm. Depending on the thickness, the observed (red) shift largely exceeds 10 GHz -*i.e.* an energy shift about 2 orders of magnitude larger than previously observed [12]- , while the apparent width of the largely distorted lineshape simultaneously increases with the decrease of the cell length, and remains an approximately constant fraction of the frequency shift (*i.e.* vW induced inhomogeneous broadening). A preliminary plot (see fig.5) of the apparent shift -as measured by evaluating the frequency of the transmission peak- as a function of the local thickness L shows an approximate $L^{-3}$ dependence, an expectation that should be in rough agreement with a rigorous theoretical modeling (for a constant relative position z/L, the shift predicted by the theory evolves indeed with $L^{-3}$, but the apparent shift of the overall lineshape is more difficult to evaluate). However, these results can be affected by the pumping process into the intermediate states - spatial inhomogeneities of the pumping through absorption, possible induced saturation or *a.c.* Stark effects- , or by the high pressure required for these experiments, implying broadening and shift.

For this reason, we have just turned to a series of systematic experiments for a limited number of cell thickness (presently in the range 40-130 nm) with investigation of the pressure effects, and control of the effects of the pumping light. When the experimental conditions ensure that the pump absorption remains weak, with the lineshapes remaining independent of the pumping power, the pumping in the $6P_{3/2}$ state can be expected to be nearly thermal.

Hence, it appears reasonable, as it had been demonstrated in the case of SR spectroscopy on another fine-structure component of the Cs 6P-6D transition [3], to extend the modelling used previously for a resonance transition to transitions between excited states, taking into account a much stronger vW interaction. On this basis, it is possible to extract an acceptable range of vW strengths from each individual spectrum, recorded for various temperatures (*i.e.* atomic densities), and to check the consistency of the extracted widths with a pressure broadening and eventual pressure-induced shift (with the vW shift and its induced inhomogenous broadening being clearly dominant, at least for small thickness experiments, the pressure shift can be most often assumed to be negligible). As a major difference with SR spectroscopy, the shift of the peak position as recorded with nanocell spectra seems to impose strict limits to the range of acceptable vW strength, while the uncertainty affecting the transition width remains much larger (conversely, in SR spectroscopy, it is often relatively easy to extract an optical width, with a large remaining uncertainty for the vW strength [3]). In addition, the experiments with nanocells permit the simultaneous recordings of the transmission and reflection lineshapes, hence providing complementary information with no need of extra-fitting parameters. This obviously imposes tight constraints to the acceptable fits. Very preliminary fits have already been obtained for experiments performed at 65 nm (see fig. 6), that provides an encouraging test of the validity of the model. One has also experimentally noted that, in spite of strong differences in reflection and transmission lineshapes, the "apparent" shifts (as defined with the peak values in transmission, or with the zeros in reflection) appear to be nearly equal for transmission and reflection lineshape, although they result [5] from a very different combination of absorption-related and dispersion-related lineshapes.

Let us also mention that in the course of these systematic experiments with nanocells, it has appeared as a crucial requirement to test the reproducibility of the recorded spectra, notably because any impurity and/or charge sticking onto the windows is susceptible to induce dramatic effects on the atomic energy levels. For our present experiments performed on a nanocell with YAG windows, lineshape reproducibility has been clearly verified when comparing different cell spots of equal thickness, as long as this thickness is 65 nm or more. The comparison of "50 nm" spots show some variations in the lineshapes, although several spots appear able to provide nearly identical responses that are selected as the "valid" response for 50 nm (see fig. 4). For different 40 nm spots, no effective reproducibility has been found. Rather, the apparent shift, relatively to the free-atom resonance, varies from spot to spot, by a factor that anyhow remains smaller than 2 in all cases. In spite of the accuracy of the thickness measurement, that apparently reaches in some cases 1-2 nm, the measurement intrinsically provides a thickness *averaged* on the optical spot size (typical beam diameter is 100-200 μm). Hence, the non-measured local irregularities affecting the window profiles (*i.e.* roughness, over various characteristic lengths), induce local fluctuations in the atom-surface distance, and are expected to produce dramatic effects on the averaged atomic spectra, especially for the smaller distances, with respect to a potential essentially sensitive to the *inverse cube* of the distance.

## 3. Resonant atom-dielectric coupling in nanocells

The theoretical model mentioned in section 2 relies on an electrostatic description of the vW interaction, with a multiple image approach to take into account the two neighbouring interacting surfaces. Such a vW description, that leads to an overall spatial dependence of the vW interaction that follows a special function (the transcendental Lerch function [14]), is obviously more precise than the simple addition of the individual potential exerted by each of the walls, even if the overall spectral predictions are most often not very different.

This electrostatic approach applies only as long as the resonant couplings between the dielectric surface and the relevant virtual atomic transitions are not considered. Indeed, the vW atom-surface interaction is a dipole-dipole interaction, originating from the coupling between the atom dipole fluctuations (to be expanded along the virtual electric dipole transitions) and the correlated fluctuations induced in the surface. Resonances in the electromagnetic coupling between a surface polariton mode and an atom fluctuation (in virtual emission), have been shown to lead to giant and possibly repulsive interaction [3,15], along with an increased decay rate from the considered excited state [16], as governed by a surface response $(\varepsilon-1)/(\varepsilon+1)$, with $\varepsilon$ the *complex* dielectric permittivity. In particular, the 12.15μm virtual emission of $Cs(6D_{3/2})$ towards $(7P_{1/2})$ makes a sapphire surface strongly repulsive for a $Cs(6D_{3/2})$ level, while a YAG surface is expected to exert only a weak repulsion [3,17]. Conversely, the $Cs(6D_{5/2})$ level, with its main virtual emission at 14.6 μm, is expected to remain attracted by a surface such as sapphire or YAG. This predicted difference between the two fine-structure components of the Cs(6D) level could not be investigated before in SR spectroscopy due to a lack of adequate sources. With the broad tuneability of our 917-921 nm source, one may naively expect to evidence these predicted differences in our present nanocell spectroscopy experiments. Actually, the situation appears to be very different for a nanocell, as due to its two coupled neighbouring walls.

Until now, our experiments performed at small thickness of a YAG nanocell have shown no sensitive differences between the two fine-structure components, but for a smaller amplitude (even when compared to the theoretical free-space ratio) for the 921 nm line $(6D_{3/2})$, and a width about twice larger (fig.4). In particular, no simple repulsive behaviour seems to be observed for the $Cs(6D_{3/2})$. In view of the recent theoretical developments [18], this does not appear very much as a surprise. Indeed, in a nanocell, the surface resonances that would be responsible for a possible enhancement of the atom-surface interaction in a half free-space, actually couple and shift *via* evanescent mode overlapping. Hence, the fluctuating atom dipole interacts with all the modes, symmetric and antisymmetric [19], of a waveguide, as appearing in an electrostatic description. A major result (fig. 7) is that for a given atomic virtual transition, the resonant coupling with the waveguide can lead to opposite behaviours between a position close to one of the walls (*e.g.* a red-shift), and a more central position (*e.g.* a blue-shift appears where an attractive potential is minimal). This behaviour cannot even be predicted from the resonant behaviour of a single wall in the limit where the atom is close to one of the walls. Indeed, for a nanocell, the waveguide thickness remains very small relatively to the IR wavelengths of the interacting atom, so that the electromagnetic field mode stucture is modified, implying modifications for the dispersive response of the atom-dielectric coupling. Besides, if the considered atom is prepared in a special manner (*e.g.* Zeeman polarised), one may become able to discriminate between the respective fluctuations of the parallel and perpendicular atom dipole, predicted to exhibit different and even opposite (repulsion *vs.* attraction) behaviours. More generally, it would be an interesting issue, whose solution is in progress, to know to which extent the resonant coupling of an atom located between two close surfaces will resemble the potential for the same atom embedded in the bulk of a similar material.

**4. Atomic collisions under confinement**

Spectroscopy close to an interface, notably SR spectroscopy, has been for a long time a choice method to evaluate collisional broadening and shift in an optically dense gas. A notable advantage of these methods is that the information is obtained close to the resonance line-centre [20], and not in the far spectral wings, as it has to be the case with standard volume experiments. However, the implicit assumption is that the vicinity with the surface has not modified the collisional features themselves. This means on the one hand that the atom dynamics -*i.e.* the velocity distribution and the population distribution- is assumed to be identical in the gas bulk and close to the surface, and on the other hand that the intrinsic mechanism of interaction between an atom and its perturber remains unaffected. For these reasons, we have started to investigate the collisional effects in nanocell spectroscopy, in the relatively elementary case of a resonant transition.

For given conditions in a nanocell, the overall pressure broadening and shift can be extracted once the Fabry-Perot effects and surface interaction have been analyzed. Assuming that the local vapour density is the same whatever the local thickness may be, it becomes also possible to analyze the significance of eventual differences appearing for various cell thicknesses. Moreover, if it is assumed that the equilibrium atomic density is the same in a nanocell and in a macroscopic vapour, varying the nanocell temperature should yield the pressure dependence (broadening and shift) of the resonances, usually expected to be linear in a large regime of density. Alternately, the amplitude of the spectroscopic signal itself is a valid indicator of the number of active atoms, yielding complementary data on the "vapour" equilibrium inside the nanocell.

Until now, our detailed experiments, performed on the well-resolved Cs $D_1$ line, have been limited to the comparison of a $\lambda/2$ and $\lambda/4$ thickness, while systematic experiments for shorter thickness, including a variation of the temperature, are planned for the near future. The frequency resolution allowed by such cell thicknesses is sufficient to discriminate the pressure behaviour of the individual hyperfine components -exhibiting considerable differences so far as the pressure-induced shift is concerned- and should enable a comparison with independent measurements through SR spectroscopy [21]. (Recently, a custom cell has been fabricated, designed for the simultaneous recording of a SR signal and of a transmission signal in nanometric regions, that should ensure similar experimental conditions for these two spectra, with no need of a comparison involving relatively inaccurate temperature measurements). As shown in section 2, preliminary theoretical fittings seem slightly improved when the vW interaction is included for the λ/4 thickness (the difference is not observable for a λ/2 cell). Remarkably, and although the extrapolated pressure shift is strongly modified when taking into account the vW shift, there appears (see fig. 8) a persisting difference in the pressure shifts for a λ/2 and λ/4 thickness, more important than the slight difference occurring in the pressure broadening once the vW interaction is taken into account. The actual strength of the vW interaction is not only derived from the optimization of the fitting spectra, but also from its ability to cancel the residual shift at null pressure, between the extrapolated resonances in the nanocell and in a macroscopic cell. However, more precise investigations are needed, because the widths become identical for λ/2 and λ/4 when the (tiny) vW interaction is neglected in the fitting process, and because of the limited number of pressure conditions that were explored for λ/4. Indeed, including a vW interaction in the fitting model shifts the estimated resonance centre, at the expense of a correlated modification of the evaluated optical width. Hence, the significance of the observed small apparent shift between the extrapolated frequency at zero-pressure and the free-space resonance, should be confirmed before ascertaining an evidence of a different pressure behaviour for λ/2 and λ/4 cells.

This possibility of a thickness-dependent collisional behaviour is worth discussing from a general point of view. An atom-atom collision, in its principle, is not very different from an atom-surface collision, especially at long distance. In particular, the atom-atom vW interaction is nothing else than the dipole-dipole coupling between the fluctuating atom-dipole, and the correlated induced (fluctuating) dipole in the atomic perturber. As long as an electromagnetic boundary lies in the vicinity of the two interacting atoms, the correlation between the fluctuations seen by the atom and its perturber can follow a variety of paths, from the direct propagation, to paths involving reflection(s) on the boundary [22]. Physically, these additional contributions can be understood as equivalent to a collision process where at least one of the atom is replaced by its electric image. In the principle, these contributions have to be considered when the distance to the reflecting plane remains smaller than the wavelength of the relevant (virtual) atomic transitions. To our knowledge, no evaluation of this confinement effect has ever been performed in the situation of a confined vapour, with the required averaging over atomic positions and velocities. Anyhow, in spite of the fact that these additional terms always exist for a nanocell (the distance to the surface being much smaller than the wavelength of the relevant virtual transitions), the dominant contribution to atom-atom collisions is expected to appear only at much shorter distances, related to the interatomic interaction potential (scaling as $r^{-6}$ - or $r^{-3}$ for a resonant collision-, with r the interatomic distance). This would imply that the confinement effect should appear only at distances much smaller than a wavelength, comparable to the one associated to the (square-rooted) collision cross-section -on the order of $10^{-12}$cm² for a resonant collision- , a distance that remains anyhow within the reach of nanocell technology.

## 5. Conclusion

It is amazing to see how high resolution techniques, employed both in the experiments and in the theoretical interpretation, are still applicable to uncooled atoms at such small distances from a macroscopic object. Also remarkable is the fact that, in spite of a strong 1D-confinement, the vapour behaves inside the nanocell according to the behaviour expected for a macroscopic vapour. Extensions of our investigations to the physics of resonant gases under strong multi-dimensional confinement, that can be approached with systems such as holey fibres or porous media [22], should offer renewed surprises, of interest for both fundamental and applied physics.

**Acknowledgments**

This work has been partially supported by European contract HPRN-CT-2002-00304, as part of the FASTNet programme. We also acknowledge the French-Uruguayan ECOS programme (U00-E03), and the specific supports of CNRS for French-Armenian cooperation (project 12856), and of Université Paris13 for scientific cooperation with Russia and C.I.S.

**Figure captions**

Figure 1 : Experimental transmission lineshapes as observed on $^{87}$Rb D$_2$ line (wavelength scan F = 2 → F' = 3,2,1) for various nanocell thickness (multiples of λ/4) at a non saturating power (~0.4 mW/cm²). The vertical scale has been normalized with respect to the nonresonant Fabry-Perot transmission. The narrow structures appearing for λ/2, and with a smaller contrast, for 3λ/2, are the signature of the coherent Dicke narrowing.

Figure 2 : Calculated peak-to-peak amplitude of the narrow structure in FM transmission, as a function of the cell length. The black points are for half-integer cell lengths (expressed in number of wavelengths), the open circles for integer lengths. The decrease of the Dicke revival peaks is exponential with a decoherence length $l_c$ proportional to the dipole mean free path, $l_c \sim 1.15\ u/\gamma$ (γ: the optical relaxation, u the thermal velocity). As due to the Doppler-broadened part of the signal that grows monotonically with the cell length, and that is not included here, the contrast between the Dicke revival peak and the total signal decreases even faster. The asymptotic nonzero limit corresponds to our initial observation of the Dicke narrowing in a microcell [1]. The Doppler/homogeneous width ratio is given by ku/γ = 25 (a); or by ku/γ = 50 (b). In the Doppler limit ku/γ →∞ , no narrow structure is found at L= nλ, yielding a zero amplitude for the FM signal (signal derivative).

Figure 3 : Reflection spectrum for the Cs D$_1$ line (F=3→ F'=3,4) for a 223 nm (λ/4) thickness. The experimental data (grey) is fitted with a theoretical lineshapes taking into account the vW attraction (black continuous line) between the two walls, or with a less satisfying model that neglects the vW interaction (white dots). The fit is here adjusted for a vW interaction governed by a coefficient ~1.8 kHz.µm$^3$ for a single YAG wall.

Figure 4 : A comparison between the transmission spectra observed respectively on the 6P$_{3/2}$-6D$_{5/2}$ (917nm, black points) and 6P$_{3/2}$-6D$_{3/2}$ (921nm, grey triangles) transitions in a 50 nm-thick Cs cell. The horizontal axis indicates the respective detuning -relatively to the free-space resonances-, the vertical one is the relative (negative) change in transmission. The pumping on the D$_2$ line induces here only negligible effects, and the results, shown here at T=240°C, (Cs density ~ 5.10$^{15}$ at.cm$^3$ ) are unchanged for a lower Cs pressure. The spectra are here

discontinuous because the laser frequency tuning is only a coarse one, measured with an accuracy close to 100 MHz.

figure 5 : A preliminary plot of the apparent frequency-shift between the transmission peak and the free-space resonances on the $6P_{3/2}$-$6D_{5/2}$ line (917 nm) and on the $6P_{3/2}$-$6D_{3/2}$ line (921 nm) as a function of the nanocell thickness L. These preliminary data, recorded with a YAG nanocell, have not been corrected for possible shifts induced by the strong atomic density (T = 300°C) or by the strong pump intensity populating Cs($6P_{3/2}$), and does not take into account the shape asymmetry attached to the vW interaction. Anyhow, the basic trend of a $L^{-3}$ behavior (continuous fitting lines) is clearly visible, with a shift following $\mathbb{C}L^{-3}$, where $\mathbb{C} \sim -440$ kHz.µm$^3$ for the 917 nm line, and $\mathbb{C} \sim -300$ kHz.µm$^3$ for the 921 nm line. This is in spite of the limited frequency resolution of the laser (~ 500 MHz).

figure 6 : Experimental reflection (a) and transmission (b) spectra as recorded on the Cs $6P_{3/2}$-$6D_{5/2}$ line (917 nm) for a 65 nm cell thickness -with YAG windows-, with examples of fits that include the vW interaction. The pressure shift is experimentally confirmed to be negligible. The fits are unsensitive to the fitting width (here taken to be 200 MHz), and the model assumes that the $D_2$ line pumping induces a "thermal" population in the $6P_{3/2}$ level (the quality of the fitting has been improved by taking a relatively small temperature of 200K). The full line assumes a 9 kHz.µm$^3$ vW interaction (for a single wall), the dotted line a 12 kHz.µm$^3$, the dashed line a 6 kHz.µm$^3$ interaction. The horizontal scale indicates the laser frequency detuning (with a ~100 MHz accuracy) relatively to the free-space resonance.

figure 7 : Resonant contribution to the potential exerted onto an atom located between two walls (at the relative distance x = z/L). Calculations are here for the parallel dipole contribution. In (a), a 3D representation of the potential curves (vertically truncated) is provided as a function of the relative frequency $\Delta_{at}$ ($\Delta_{at} = \omega_{at}/\omega_s$) of the atom virtual transition (assumed to be a single one, at $\omega_{at}$) relatively to the surface resonance $\omega_s$. This dispersive resonance (whose relative width is here assumed to be 0.04) implies an atom energy shift for a single wall that goes from positive to negative values when the relative atomic frequency $\Delta_{at}$ ($\Delta_{at} = \omega_{at}/\omega_s$) goes across the surface resonance ($\Delta_{at}=1$). In (b), the potential curves appear to be surface-attractive ($\Delta_{at}=1.0308$) or surface-repulsive ($\Delta_{at}=1.0312$), irrespectively of the sign

of the energy-shift (*i.e.* the attraction is here compatible with a positive-shift at the centre for $\Delta_{at}$=1.0308).

Figure 8 : Width (a) and shift (b,c) as a function of the Cs pressure, as obtained from a fitting of the transmission lineshapes for the F=4 →F'=4 component of the $D_1$ line. In (a), the open squares are for a λ/2 thickness (identical widths with and without van der Waals), the open circles and stars for a λ/4 thickness, respectively with and without the vW interaction taken into account ; in (b) and (c), respectively standing for a λ/2 and a λ/4 cell thickness, the full circles are for a model with the vW interaction taken into account, the open squares are for a model neglecting the vW interaction. The vW interaction strength is taken to be 1.8kHz.µm$^3$ for a single wall.

Hamdi et al.,    Laser Spectroscopy with nanometric gas cells,    Figure 1

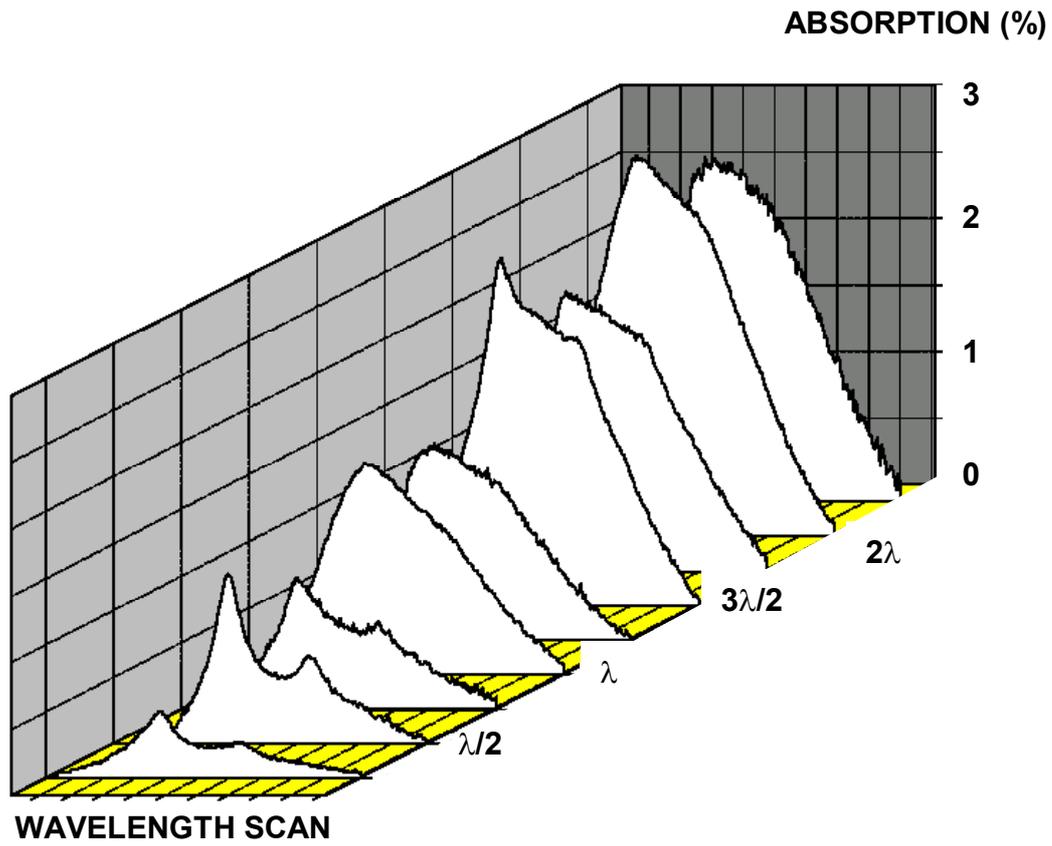

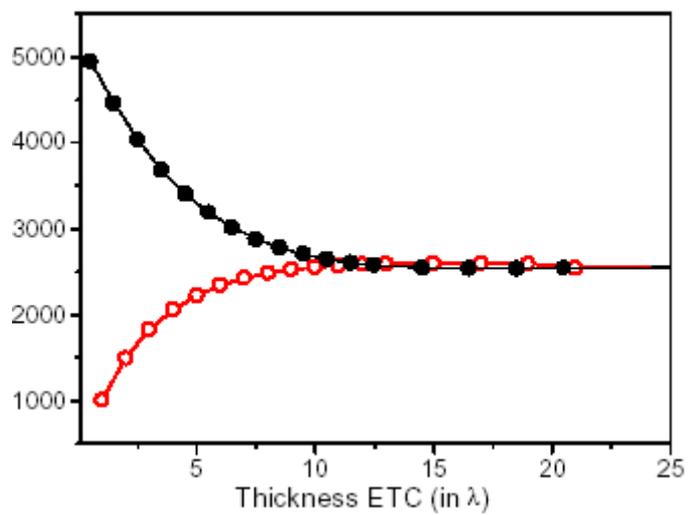

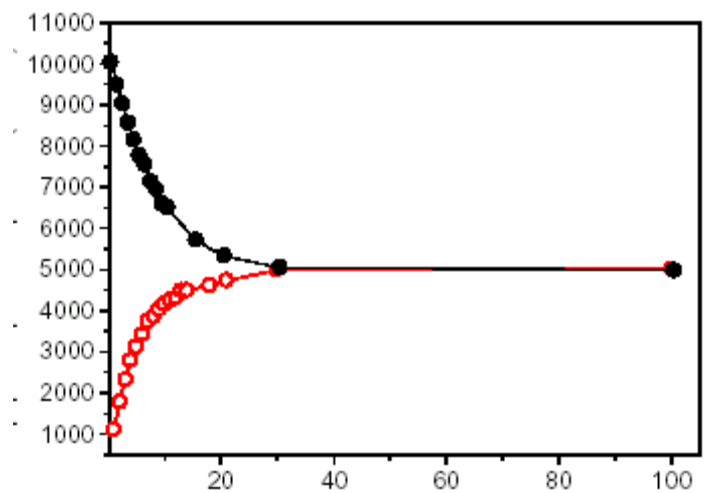

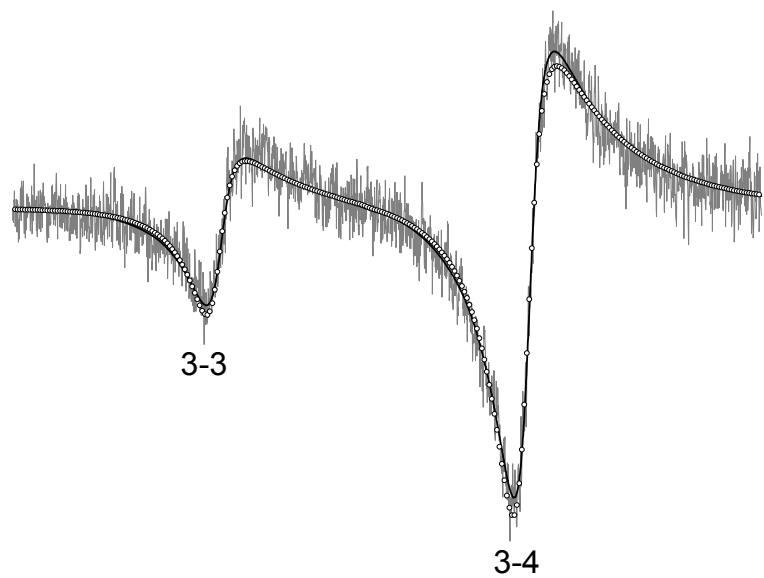

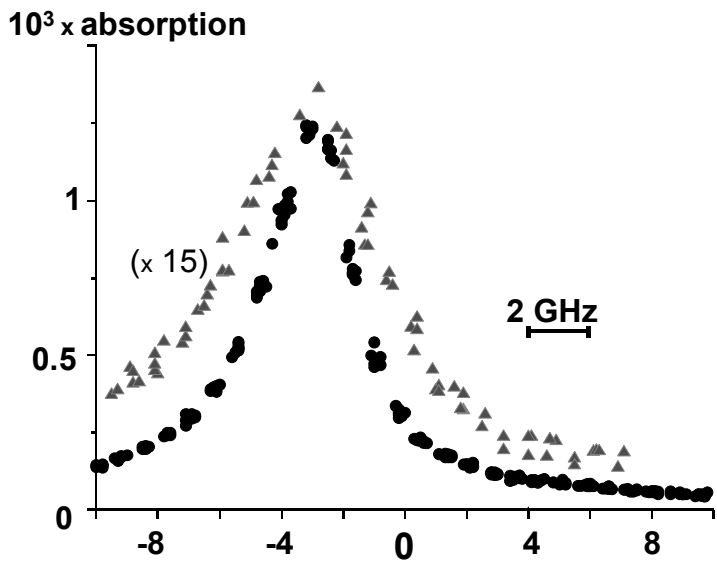

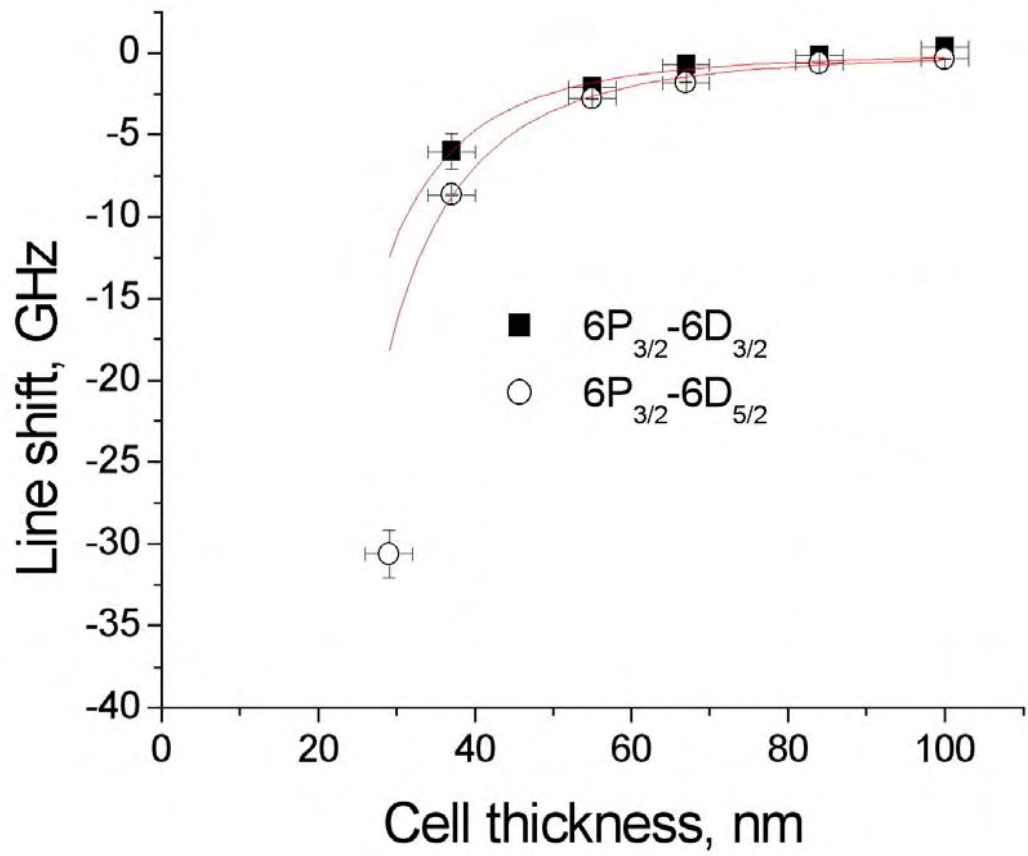

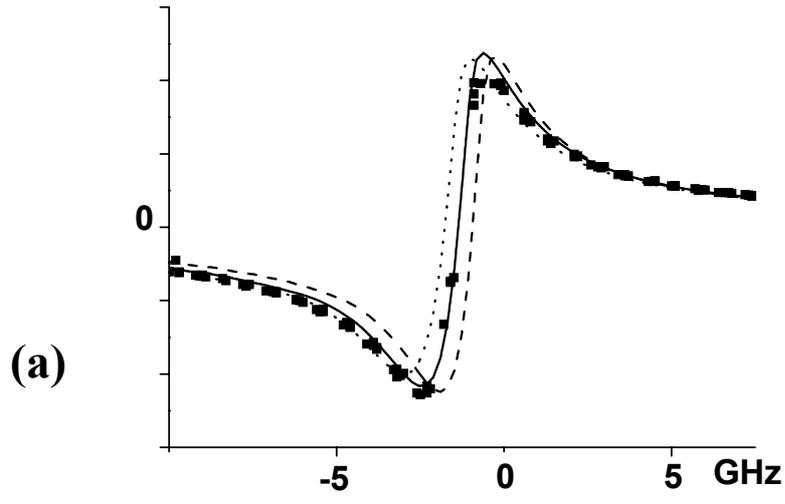

**(a)**

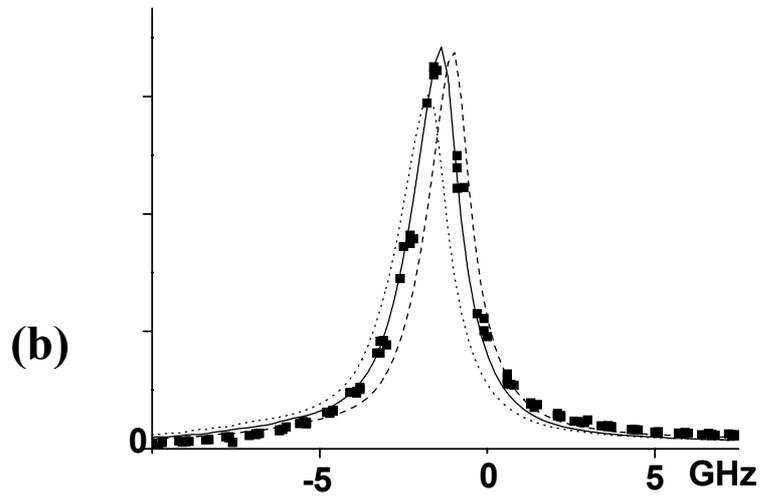

**(b)**

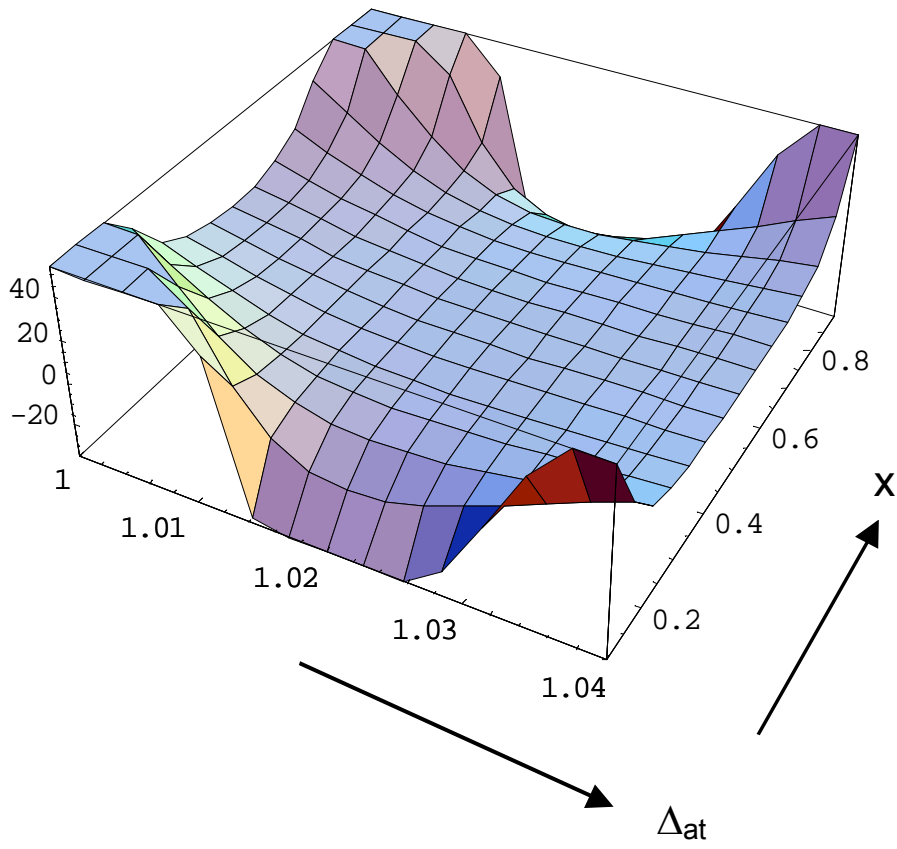

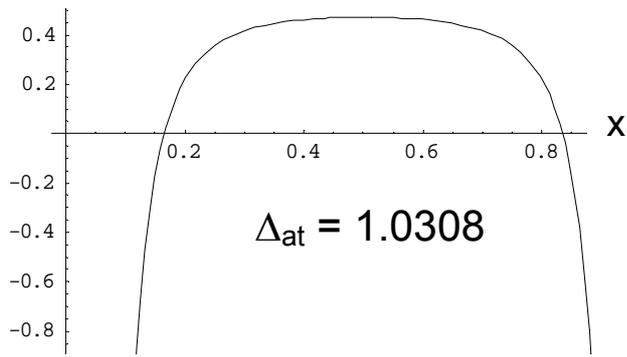
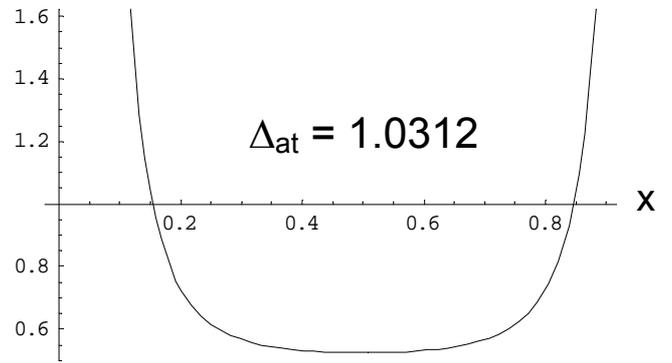

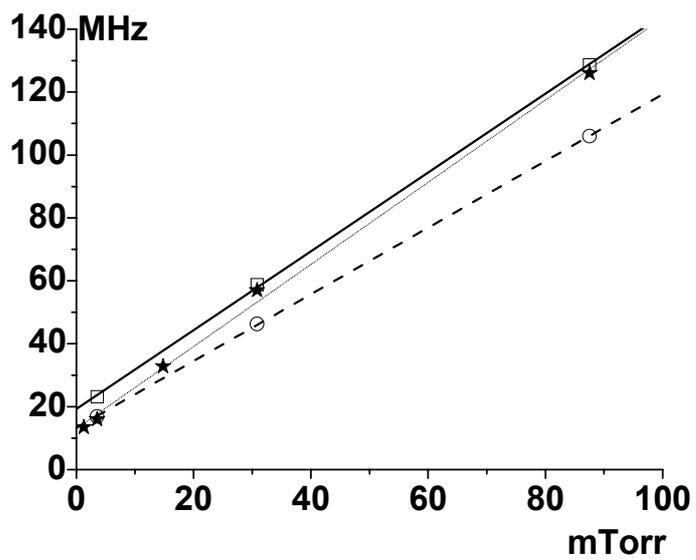

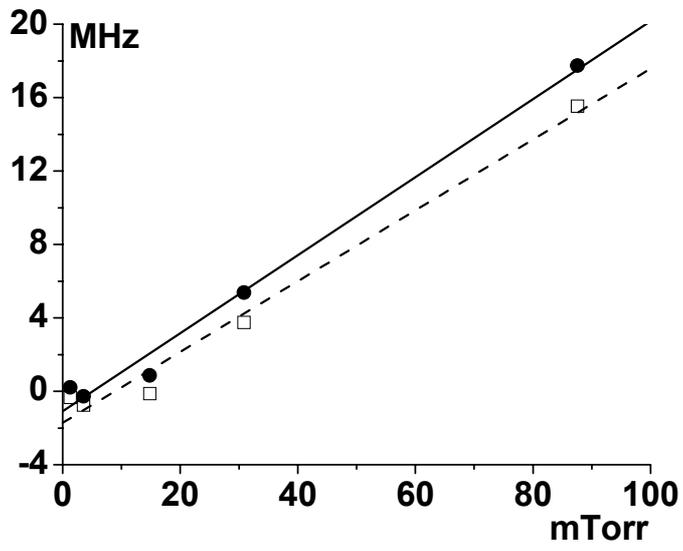

(b)

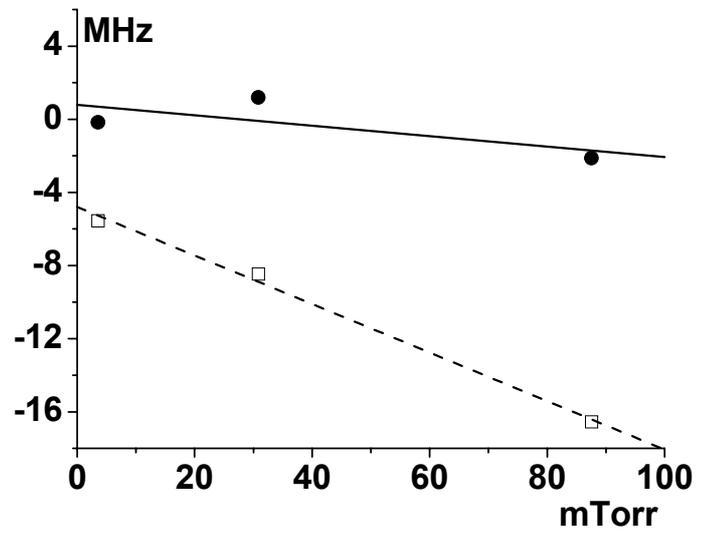

(c)